\documentclass[final,3p,times,twocolumn]{elsarticle}

\usepackage{amssymb,amsmath,amsthm,bm}
\usepackage{geometry}
\usepackage{url}

\renewcommand{\vec}[1]{\mathbf{#1}}

\journal{Journal of Information Security and Applications}
\usepackage{hyperref}

\begin{document}

\begin{frontmatter}

\title{When an attacker meets a cipher-image in 2018: A Year in Review}

\author[hnu-cn]{Chengqing Li\corref{corr}}
\ead{DrChengqingLi@gmail.com}

\author[hnu-cn]{Yun Zhang}

\author[xtu-cn]{Eric Yong Xie}

\cortext[corr]{Corresponding author.}

\address[hnu-cn]{College of Computer Science and Electronic Engineering, Hunan University, Changsha 410082, China}

\address[xtu-cn]{College of Information Engineering, Xiangtan University, Xiangtan 411105, Hunan, China}

\begin{abstract}
This paper aims to review the encountered technical contradictions when an attacker meets the cipher-images encrypted by the image encryption schemes (algorithms) proposed in 2018 from the viewpoint of an image cryptanalyst. The most representative works among them are selected and classified according to their essential structures. Almost all image cryptanalysis
works published in 2018 are surveyed due to their small number. The challenging problems on design and analysis of image encryption schemes are summarized to receive the attentions of both designers and attackers (cryptanalysts) of image encryption schemes, which may promote solving scenario-oriented image security problems with new technologies.
\end{abstract}
\end{frontmatter}

\section{Introduction}
\label{sec:intro}

Security and privacy protection of image data is an everlasting challenge in the cyberspace.
Designing image encryption schemes (algorithms, techniques, methods) and further processing in the domain of the cipher-images 
received intensive attention from researchers in the field of security, signal processing, nonlinear science, and etc \cite{cqli:IEAIE:IE18}.
From Web of Science Core Collection, we found about 1,000 publication records published in the year 2018 by inputting ``image'' and ``cryptograph* or encrypt*'' in the field of ``Topic''. Some more were published in some top conferences on nonlinear sciences and signal processing, e.g. \href{https://dblp.uni-trier.de/db/conf/iscas/}{ISCAS} and \href{https://dblp.uni-trier.de/db/conf/icmcs/}{ICME}.

In cryptography, substitution-permutation network is the structure of some modern block ciphers such as AES (Advanced Encryption Standard), where an S-box (substitution-box) is used to substitute the input of the S-box by another block of bits. Following the framework of Fridrich's image encryption scheme cryptanalyzed in \cite{Cqli:Fridrich:SP2017}, a large number of image encryption schemes adopting permutation-then-diffusion without substitution-box (PTDWOS) have been proposed since 1998 \cite{chai:brownian:MTA2018,Ye:pixel:ND2018,yuhai:adaptive:SP2018},
where \emph{position permutation} can be represented as
\begin{equation}
\vec{I}^*(w(i))=\vec{I}(i),
\end{equation}
\textit{permutation vector} $\vec{W}=\{w(i)\}$ is a bijective map on $\vec{D}$, the supporting domain of the plain-image, $i\in \vec{D}$;
\emph{value diffusion} is performed via substitution function
\begin{equation}
\vec{I}'(i)=\vec{I}^*(i)\boxplus g(\vec{I}'(i-1))\boxplus h(i),
\end{equation}
$\boxplus$ denotes an arithmetic operation, $g$ is a fixed nonlinear function, and $\vec{H}=\{h(i)\}$ is a pseudo-random number sequence (PRNS).
Only small number of chaotic image encryption schemes use S-box \cite{Zhucx:analysis:Symmetry2018,Liu:box:MTA2018,Ullah:box:ND2018}. There are some papers focusing on designing S-box with chaotic maps \cite{Gonzalez:box:ND2018,Khan:box:NCA2018,Alzaidi:IA:2018,Luengas:tent:MTA2018}.
In \cite{Zhucx:analysis:Symmetry2018,Diab:permutation:SP2018,Wu:permutation:SP2018,Lim:Cryptanalysis:IM2018,Wang:Cryptanalysis:SP2018,Panwar:Cryptanalysis:JEI2018,
Chen:Cryptanalysis:ND2018,Zhucx:cryptanalysis:En2018,Feng:encode:IPJ2018}, the analyzed image encryption schemes are fixed to withstand the proposed attacking methods. Due to the seemingly similar properties of a chaotic system and a secure
pseudo-random number generator (PRNG), a large number of image encryption schemes use various chaotic maps to generate PRNS, which is used to control combination of some basic encryption operations, e.g. Logistic map \cite{Elsadany:logistic:AMC2018,Huang:wave:MTA2018}, Tent map \cite{Luengas:tent:MTA2018,Ahmad:tent:NCA2018}, Cat map \cite{Hua:cat:TC2018}, PWLCM (piecewise linear chaotic map) \cite{Luo:chaos:MTA2018,Zhang:chaos:MTA2018}, and coupled map lattice \cite{Kumar:model:JISA2018}. Rigorously, only the encryption schemes using special chaos methodologies can be classified as chaos-based encryption schemes, e.g. chaos synchronization \cite{Lakshmanan:Synch:TNNLS2018,Hu:synchronization:TII2018}.
As the opposite of image cryptography, some impressive image cryptanalysis works were developed in 2018 \cite{Diab:hyper:SP2018,Lim:Cryptanalysis:MTA2018,Ahmad:Cryptanalysis:Symmetry2018,Luo:hardware:IJBC:2018,Fouad:stream:SP2018,ozkaynak:review:ND18}. Especially, all image encryption schemes published in the journal Nonlinear Dynamics before 2018 are critically reviewed in \cite{ozkaynak:review:ND18}.


This paper selected and classified the representative methods of protecting and disclosing image data proposed in 2018. Some challenges existing in the two sides are
concluded from the perspective of a senior cryptanalyst. For example, only a few image encryption schemes focus on real specific applications, e.g. wireless communication
\cite{Helmy:cube:MTA2018}, surveillance \cite{Muhammad:surveillance:TII2018}. This paper aims to let designers and cryptanalysts of image encryption (and privacy protection)
schemes glimpse the annual progress on confrontation between attackers and protectors of image data.

The remainder of the paper is organized as follows. Section~\ref{crypto} reviews the design of image encryption schemes proposed in 2018. Then, Sec.~\ref{cryptanalysis} presents a survey on cryptanalytic works on image security given in the last year. Section~\ref{challenge} summarizes the challenges on the two sides on image security. The last section concludes the paper.

\section{Survey on image cryptography in 2018}
\label{crypto}


\subsection{Image encryption schemes based on chaos}

According to essential structure and used methodologies, the encryption schemes using chaotic system are coarsely classified in the following five sub-sections.

\subsubsection{Randomness-oriented chaos enhancement}

To counteract dynamics degradation of chaotic maps in digital computer and enhance
randomness of PRNS generated by iterating a chaotic system \cite{lu:iteratE:en2018}, various strategies are
proposed, e.g. cascading two existing chaotic maps \cite{Hua:Sine:TIE2018};
iteratively expanding a parametric 2-D Cat matrix to any higher dimension \cite{Hua:cat:TC2018};
arbitrarily combining six basic nonlinear operations \cite{Hua:model:TCASI2018}; anti-control 
\cite{Yusm:video:TCASVT2018}; constructing hyper-chaotic system \cite{Bouslehi:hyper:MTA2018}. In \cite{Hua:Sine:TIE2018}, Hua et al. design a sine-transform-based chaotic system of generating
one-dimensional (1-D) chaotic maps by performing a sine transform to the combination of the outputs of two maps. In \cite{Hua:sine:SP2018}, Hua et al. design a two-dimensional (2-D) Logistic-Sine-coupling map and use it as a source of PRNS controlling basic encryption operations of an image encryption scheme, whose structure follows PTDWOS. The same strategy is used in \cite{Gayathri:chaotic:MTA2018}, where a spatiotemporal chaotic system is coupled with Tent-Sine system to generate PRNS. In \cite{Hua:scrambling:SP2018}, an encryption scheme of protecting medical images with two rounds of position permutation and diffusion, where Logistic-Sine system is used as PRNG.

In \cite{Yang:logistic:SP2018}, the periodicity of Logistic map and its variants
over finite field $\mathbb{Z}_{3^e}$ is analyzed. Based on comprehensive
analysis of randomness of the PRNS generated by iterating
a map, one variant of Logistic map is selected in \cite{Yang:logistic:SP2018} as a
PRNG. In \cite{Zhu:effect:MTA2018}, a 6-D discrete chaotic system
is constructed with some sine and cosine functions, which
is then used as a source of PRNS used in a stream cipher.
As analyzed in \cite{Elsadany:logistic:AMC2018}, the coupled logistic map can demonstrate
complex bifurcation dynamics in a continuous domain. Actually,
the real structure of the final chaotic systems obtained
with any above way in a digital computer should be
analyzed as \cite{cqli:network:TCASI2019} with the methodology of state-mapping network.

\subsubsection{Design of single round PTDWOS}

Among the encryption schemes composing of only one round of basic operations, only \cite{Luengas:tent:MTA2018} adopt
S-box, where tent map is used as PRNG to control some basic encryption operations on image: modular addition, S-box and binary shift operation.

In \cite{wangch:search:IJBC2018}, 1-round PTDWOS is proposed, where
the algorithm on traversing a graph with the strategy of breadth-first search is used to 
implement position permutation, and a hyper-chaotic system is used as PRNG to realize the
second permutation and diffusion with XOR and modular addition. In\cite{Liu:block:IETSP2018}, an encryption function on image blocks with variable sizes is proposed, which is composed of position permutation and modulo addition. The two basic operations are controlled by Arnold map and integer Logistic map, respectively. In \cite{Mondal:diffusion:MTA2018}, the proposed image encryption scheme performs only one round of 
position permutation and bitwise exclusive OR (XOR) operation, which are both controlled by PRNS generated by 2-D Baker's map. The critical frame of the surveillance video for IoT systems is
encrypted the two steps in \cite{Muhammad:surveillance:TII2018}, where 2-D logistic-adjusted-sine
map (LASM) is used as the source of PRNS. The dependence of PRNS on the plain-image can be canceled 
with a non-negligible probability due to the commutative principle of summation and quantization error of calculating function in a finite-precision computer. In \cite{Liu:medical:MTA2018}, a simple fourth-order chaotic system is constructed as PRNG to control the XOR operation and position permutation on a medical image. As only one round of the two basic operations are used, the encryption schemes designed in \cite{Liu:medical:MTA2018,Muhammad:surveillance:TII2018} are both vulnerable to chosen-plaintext attack. 

In \cite{Ping:substitution:Neurocom2018}, Ping et al. encrypt two adjacent pixels with position permutation controlled by 2-D H\'enon map and modulo addition. In \cite{Ping:ca:SP2018}, Ping et al. use 2-D Logistic-adjusted-Sine map to implement the permutation part and 2-D Cellular Automata (CA) as the source of PRNS to control the part on modulo addition. In \cite{Sahari:PRNS:ND2018}, a PRNG based on a 3-D chaotic map is designed. The framework of the image encryption scheme is the same as that of \cite{Ping:ca:SP2018}. In \cite{Ping:block:IA2018}, Ping et al. propose a method of digit-level permutation, which can change the histogram of the permuted image. In \cite{Huang:wave:MTA2018}, 2-D Logistic map is used to generate PRNG to control combination of position permutation and diffusion with XOR. 

In \cite{Liz:hyper:ND2018}, nine special pixels in the plain-image are selected to generate initial conditions of Lorenz system with Secure Hash Algorithm SHA-256. However, the sensitivity mechanism can be easily canceled as only nine selected pixels have no influence on other ones. In \cite{Luo:chaos:MTA2018}, a self-adapting colour image encryption scheme based on chaotic maps and the interactions among three channels is proposed. Due to the commutative property of modulo addition, the sensitivity on the plain-image can be easily canceled as well.

\subsubsection{Design of multiple round PTDWOS}

Reference \cite{Sheela:henon:MTA2018} proposes an image encryption scheme based on two rounds of position permutation and XOR operation, which are controlled by PRNSs generated by 2-D 
modified Henon map and Sine map, respectively. Reference
\cite{Rajagopalan:communication:MTA2018} adopts the same encryption structure as \cite{Sheela:henon:MTA2018} with
hardware implementation, where Lorenz chaotic system,
L\"u's chaotic system and CA are used as the
source of PRNS instead. In \cite{Teng:permutation:MTA2018}, multiple rounds of PTDWOS
are implemented in the level of bit plane, where Skew
Tent map is used as source of PRNS. In \cite{Ye:pixel:ND2018}, 2-D Sine-Logistic modulation map (2D-SLMM) is used to control the two basic part of a scheme following
PTDWOS, where modular addition with a fixed matrix is
adopted to enhance the diffusion effect. In \cite{Cao:hyper:SP2018}, a bit-level
image encryption algorithm is designed by operating multiple
rounds of bitwise shift operation and XOR in two directions,
where a hyperchaotic map is constructed as the source of PRNG by cascading sine map and Logistic map.

Following the basic structure of AES, a robust image symmetric cryptosystem
is proposed in \cite{Becheikh:Risc:MTA2018}, where Arnold's cat map is used to generate a permutation matrix permuting the entire plain-image.
In \cite{Ghebleh:plcm:MTA2018}, an image encryption algorithm based on 1-D PWLCM and least squares approximation is proposed, where PWLCM is used to determine the permutation relationship among the rows and columns of the plain-image and the intermediate matrixes. The method based on least squares approximation is adopted to generate PRNS to mask the permuted intermediate matrixes. In \cite{Zhang:chaos:MTA2018}, PWLCM is used as PRNG to control the combination of two rounds of position permutation and modular addition.
In \cite{Wang:CA:Neuro2018}, a 2-D partitioned CA is designed to generate PRNS satisfying the global Strict Avalanche Criterion, which is used to determine the combination of some basic operations including substitution with an S-box.

\subsubsection{Encrypting multiple plain-images simultaneously}

In \cite{chai:brownian:MTA2018}, a double colour image encryptionDCIE) scheme using 3-D Brownian motion is presented, where two colour images are encrypted simultaneously. DCIE follows the structure of PTDWOS and uses 3-D Brownian motion, a 3-D autonomous chaotic system and 2D-LASM as PRNG to control its basic components. In \cite{Shao:phase:MTA2018}, another DCIE using Gyrator transform and H\'{e}non map is proposed. In \cite{Hanis:compression:MTA2018}, the four least significant bit-planes of two images are encrypted by position permutation and value confusion, which are controlled by Logistic map and CA, respectively. In \cite{Yu:transform:MTA2018}, four plain-images are encrypted simultaneously using 2D-LASM. In \cite{Wang:tensor:MTA2018}, a sequence of images are simultaneously encrypted and compressed using Tensor Compressive Sensing, which is controlled by PRNS generated by a 3-D discrete Lorenz system.
In \cite{Shao:Multiple:MTA2018}, a multiple optical colour image encryption scheme based on phase retrieval in quaternion gyrator domain is proposed, nonlinear quaternion correlation is developed to perform authentication.

In \cite{Kaur:Beta:IJBC2018}, an image encryption scheme using beta chaotic 
map, NSCT (Nonsubsampled Contourlet Transform), and
GA (genetic algorithm) is proposed. The plain-image is decomposed
into three subbands. Then, GA is used to select the optimized
initial parameters of the beta chaotic map for a multi-objective
fitness function. Finally, the inverse of NSCT
is applied on the encrypted subbands with modulo addition
to produce the cipher-image. In \cite{Noshadian:Optimizing:MTA2018}, teaching learning
based optimization method and gravitational search algorithm
are used to optimize selection of the parameters.
However, such optimization algorithms cost additional unknown
computational load. In \cite{Mozaffari:Parallel:MTA2018}, GA is used to perform
parallel permutation and substitution on multiple bitplanes
of the plain-image.

\subsubsection{Encrypting images with compressing techniques}

In \cite{Huang:diffusion:SP2018}, an image encryption scheme based on compressive sensing and chaotic map is proposed. The 2D-SLMM is used in the three basic parts of the scheme: constructing measurement matrix in compression operation; generating PRNG for diffusion operations; producing permutation matrix. In \cite{liaoxf:encrypt:MTA2018}, a plain-image is first decomposed by wavelet packet transform, whose coefficients are classified in terms of the average value and Shannon entropy. The 
coefficients containing principal energy are encrypted by position permutation, modular subtraction and 
XOR, which are controlled by Logistic map, Chen system and Arnold map, respectively. Other nonzero coefficients are compressed by compressive sensing (CS) with a secret measurement matrix constructed from a Hadamard matrix controlled by Cat map. The encryption scheme is based on 2-D CS and fractional Fourier transform (FrFT). In \cite{liaoxf:compressive:MTA2018}, Liao et al. employ Kronecker product to generate higher dimensional measurement matrices. In \cite{Pudi:cs:TCSII2018}, compressive sensing technique is combined with a stream
cipher to simultaneously compress and encrypt image and video files, where the Measurement Matrix is generated  using a stream cipher. Only the four
most significant bits of compressed samples are encrypted, which may cause some and even all visual information of the encrypted object revealed.

\subsection{Designing image encryption schemes based on DNA encoding}

With the development of DNA computing, designing image encryption schemes based on DNA encoding received intensive attentions in the past decade \cite{Wang:dna:MTA2018}. In 2018, about twenty technical papers on the topic were published. Here, we review representative works among them.

In \cite{Maddodi:DNA:MTA2018}, a new heterogeneous chaotic neural network generator is proposed as a source of PRNS to control the three basic encryption operations on image data: pixel position permutation; DNA-based bit substitution; DNA-based bit permutation. 
In \cite{Girdhar:DNA:MTA2018}, a robust color image encryption
system using Lorenz-Rossler chaotic map and DNA encoding
is proposed, where Lorenz-Rossler chaotic map is
used as the source for producing PRNS. The three channels of
color plain-image and PRNS are transformed into the form
of DNA strands. After performing subtraction and addition
operations, the results are transformed back to the binary
form as the cipher-image. In \cite{BenSlimane:DNA:MTA2018}, an image encryption
scheme based on DNA encoding is designed, using a 2D-LASM and Single Neuron
Model as sources of PRNS. First, the plain-image is encrypted by position
permutation and XOR. Then, the intermediate image
is further encrypted in the DNA domain with XOR operation.

In \cite{Fu:DNA:IPJ2018}, a scheme encrypting optical image is designed with the same strategy as that in \cite{Maddodi:DNA:MTA2018,BenSlimane:DNA:MTA2018}. In \cite{Sun:dna:IPJ2018}, an image encryption scheme using pixel-level scrambling, bit-level scrambling, and DNA encoding is proposed, where a 5-D hyperchaotic system is used as the source of PRNS. In \cite{Wu:henon:SP2018}, 2-D H\'{e}non-Sine map is constructed as a PRNG to control the involved DNA rules, DNA operation and the position permutation.
In \cite{Wen:meaningful:NCA2018}, the saliency detection of a plain-image is detected with
a given model and encrypted with a chaos-based encryption function. Then, the encrypted
result is embedded into DCT domain of another meaningful natural image as an invisible watermark.

\subsection{Encrypting image in transform domain}

A series of encryption schemes protecting optical images
is proposed in 2018. But, most of them are simulated
with digital images due to the limit of optical devices. In \cite{Lixw:optical:OE2018}, a monospectral synthetic
aperture integral imaging system is designed to capture
2-D optical elemental images, which are then XORed
with a PRNS generated by CA and digitally encoded by Fresnel
transform. In \cite{Ren:fourier:OR2018}, the original image is first scrambled,
and then multiplied by the random-phase mask function. A
further phase-truncated discrete multiple-parameter FrFT is
implemented to realize asymmetric encryption. In \cite{Kong:image:TII2018}, a
holographic encryption scheme based on interleaving operation
of computer-generated holograms is proposed.
Using the quaternion algebra, B. Chen et al. define
a multiple-parameter fractional quaternion Fourier
transform as a general version of the conventional
multiple-parameter fractional Fourier transform \cite{Chenbj:fourier:IET2018}. Detailed experimental results demonstrate
that MPFrQFT is superior to other eight existing algorithms
on multiple metrics: key space,
key sensitivity, statistical analysis and robustness.

In \cite{Chen:hyperspectral:OLE2018}, an optical hyperspectral image cryptosystem is proposed using improved Chirikov mapping in the gyrator transform domain, the original hyperspectral image is converted into binary format and then extended into a 1-D array, then the scrambled and exchanged image is transformed using the gyrator transform. In \cite{Kumar:JMO:JMO2018}, quick response (QR) code-based non-linear technique for image encryption using shearlet transform and spiral phase transform is proposed. The input image is first converted into a QR code and then scrambled via the Arnold transform.

In \cite{zhoujt:privacy:TMCCA2018}, a DCT-domain image encryption framework is proposed with block-wise permutation and XOR operation, which is implemented with a stream cipher. In \cite{He:encryption:ITM2018}, a bitstream-based JPEG image encryption scheme with negligible size expansion is proposed, which cascades 
four operations: permuting the groups of successive DC codes encoding
the differences of quantized DC coefficient with the same sign;
swapping the two half parts of a group of
consecutive DC codes; scrambling the AC codes falling the same category;
randomly shuffling all minimum coded units.

\subsection{Signal processing in the encrypted domain}

In \cite{Huangjw:image:TIP2018}, the fast methods of implementing 
real and complex Walsh-Hadamard Transform in the cipher-images are proposed.
To further reduce the computational complexity of the homomorphic encryption operations, two parallelization strategies are given to accelerate the transform.
Using strong correlation among neighbouring pixels in a natural image, an effective high-capacity reversible data hiding scheme for encrypted images based on MSB (most significant bit) prediction is proposed in \cite{Puteaux:hiding:TIFS2018}. Interestingly, \cite{Puteaux:hiding:TIFS2018} adopts PWLCM as the source producing pseudo-random number sequence, whose randomness is seriously comprised in a digital computer \cite{cqli:network:TCASI2019}.

Reversible data hiding in encrypted image (RDH-EI) embeds data into a cipher-image in cloud meanwhile the corresponding plain-image can be perfectly reconstructed by the authorized receiver. In \cite{Chen:hiding:JVCIR2018}, two RDH-EI schemes are proposed with private-key homomorphism and public-key homomorphism, respectively. In \cite{Bhardwaj:hide:JEI2018},
an enhanced RDH-EI embedding two bits in each cipher-pixel with minimum distortion of stego-pixel is achieved through homomorphic encryption. In contrast to many data hiding schemes focusing on LSB, an efficient MSB prediction-based method for RDH-EI is proposed in \cite{Puteaux:Hide:TIFS2018}, where vacating embedding room after encryption and 
reserving embedding room before encryption are discussed, respectively.
To cope with the compressed images, \cite{zhangxp:hide:TDSC2018} proposes a separable RDH for encrypted JPEG bitstreams, where the original bitstream is losslessly reconstructed using an iterative recovery algorithm. In \cite{Xiang:hide:TCSVT2018}, a RDH-EI is designed using homomorphic multiplication and probabilistic properties of Paillier cryptosystem,
where the embedding room is reserved before encryption.

In \cite{Wang:compress:JVCIR2018}, C. Wang et al. formulate lossy compression
on low-frequency wavelet coefficients as a problem of weighted rate-distortion optimization and solve it by incorporating empirical characteristic of their distribution.
In \cite{Wang:markov:TIFS2018}, C. Wang et al. propose an iterative reconstruction
scheme for compression file of encrypted binary images
using the Markov Random Field to characterize
the corresponding plain-images in the spatial domain.
The associate Markov Random Field, decryption and decompression based
on LDPC are all represented
with the methodology of factor graph, a bipartite graph
representing the factorization of a function into some subfunctions.

\subsection{Generating cipher-images in other application scenarios}

\begin{itemize}
\item Visual secret sharing

The $(k, n)$-visual secret sharing (VSS) scheme encode a secret image
into $n$ shares, such that any $k$ shares can be superimposed to reveal the secret image to human's eyes but any $k-1$ or less ones can leak nothing.
In \cite{Shyu:cryptographic:TCASVT2018}, three efficient and flexible XOR-based $(k, n)$-VSS schemes is constructed, such that more than $k$ shares can also reveal the secret image. As for progressive $(k, n)$ secret image sharing (PSIS) scheme, it is required that $k$ to $n$ shares can reconstruct the secret image progressively.
In \cite{Liu:sharing:SPIC2018}, three PSIS schemes are constructed using XOR operation, a variant of Hamming code and modified version of shorten Hamming code, respectively.
In \cite{Yan:share:DSP2018}, an image secret sharing with three decoding options, 
lossless recovery, grayscale stacking recovery and visual previewing, is designed based on random grid-based VSS and Chinese remainder theorem.   

\item Privacy-preserving application of image data

In \cite{Yuan:privacy:TDSC2018}, Yuan et al. use similarity degree among encrypted higher-dimensional profile vectors of users’ images as a metric to search and recommend friends in social media. In \cite{Gunasinghe:authentication:TIFS2018}, a privacy-preserving biometrics-based
authentication solution is proposed to authenticate the users to remote service providers
via zero-knowledge proof of knowledge on two secrets: an identity token encoding the biometric identifier of the user's image and a secret owned by the user. In \cite{Wang:Searchable:TDSC2018}, a searchable symmetric encryption over encrypted multimedia data is proposed by considering the search criteria as a high-dimensional feature vector, which is further mapped by locality sensitive hashing. The inverted file identifier vectors are encrypted by an additive homomorphic encryption or pseudo-random position permutations.
In \cite{Xia:cloud:TCC2018}, a privacy-preserving content-based image retrieval (CBIR) scheme is proposed to allow the data owner outsource the image database and CBIR operation to the cloud server without disclosing its actual content. The proposed scheme let the cloud server can solve earth 
mover’s distance as a linear programming problem without sensitive information of the searched image.

\item Quantum image encryption

The development of quantum cryptography and quantum chaos attracts image presentation in quantum world and study on how to utilize quantum properties for protecting image data \cite{Khan:quantum:Plos2018,Ahmed:IA:2018}.
In \cite{Zhounr:quantum:QIP2018}, a quantum representation model is proposed 
to let quantum hardware encrypt any number of plain-images simultaneously. Then, the quantum circuit of quantum 3D Arnold transform is defined. Numerical simulations are given to show performance of a multi-image encryption scheme combining quantum 3D Arnold transform, quantum XOR operations and scaled Zhongtang chaotic system.

\item Permutation against attacks based on ``jigsaw puzzle solver''

Due to simplicity and obvious security enhancing effect of position permutation,
it is widely used in image encryption scheme \cite{He:encryption:ITM2018,Cqli:Scramble:IM17}. The ciphertext-only attack on
any block-wise permutation scheme can be attributed to solving a jigsaw puzzle.
Recently, Kiya et al. published a series of works on encrypting each image block with marginal influence on statistics and permuting the encrypted blocks
among the whole image. Such encryption-then-compression framework does not seriously influence the further compression and can withstand the attacks based on conventional jigsaw puzzle solvers \cite{Kiya:scrambling:ITFECCS2018}.
But, the correlation among blocks in cipher-images may still facilitate 
optimizing the attacking on the block-wise permutation schemes analyzed in \cite{Kiya:block:ITIS2018,Cqli:hierarchical:SP2016} with the advanced jigsaw puzzle solvers.
\end{itemize}

\section{Survey on image cryptanalysis in 2018}
\label{cryptanalysis}

In the year 2018, about 30 papers on image cryptanalysis were published. Among them, the most striking work is depreciating motivation of most chaos-based encryption schemes in \cite{Preishuber:motivation:TIFS2018}, refuting any superiority of chaos-based encryption schemes than several variants of AES in terms of computational effort and security performance, which may deserve argument with rigorous theoretical analyse or convincing engineering application. In \cite{Preishuber:motivation:TIFS2018}, four obviously insecure encryption schemes are subtly constructed to question credibility of the security assessment metrics used in the field of chaotic cryptography widely, e.g. correlation, entropy, histogram variance, Number of Pixel Change Rate (NPCR), Unified Average Changing Intensity (UACI), key sensitivity, robustness against differential attack, sequence test, and NIST randomness test. Considering the quantitative test results of the four special schemes
on the metrics, it is concluded that the metrics are far ineffective for security analysis, which are further qualitatively analyzed the inefficiency of the metrics in \cite{cqli:IEAIE:IE18}. Other image cryptanalysis works are separately reviewed in the following three subsections.

\subsection{Cryptanalysis of single round PTDWOS}

In \cite{cqli:autoblock:IEEEM18}, security of a block-wise image encryption scheme using an ECG signal and the plain-image to determine the used PRNG is analyzed comprehensively. As the sensitivity mechanism on the plain-image is built via a modulo addition, it can be canceled with a non-negligible probability of 1/256. Meanwhile, no permutation operation is adopted in the scheme, causing an equivalent secret key can be derived from one pair of a known plain-image and its corresponding cipher-image. As the insecurity of sole permutation analyzed in \cite{ZhangY2017InfSci},
position permutation is still a simple but efficient measure to improve the security level of the whole encryption scheme.

Four research teams independently studied security analysis of an image encryption scheme using the combination of some 1-D chaotic maps
proposed by Pak et al. in 2017, whose structure is one-round PTDWOS with an additional shift permutation.
Ignoring the fact that permutation and modulo addition are not commutative, Wang et al. simplifies Pak's scheme by cascading
two non-neighbouring permutations together and break it with a chosen-plaintext attack in \cite{Wang:Cryptanalysis:SP2018}.
The enhanced version of Pak's scheme suggested in \cite{Wang:Cryptanalysis:SP2018} is found still vulnerable against chosen-plaintext attack \cite{Panwar:Cryptanalysis:JEI2018}. Assuming a parameter seed is known, the parameter determining the shift permutation of Pak's scheme is disclosed in \cite{Chen:Cryptanalysis:ND2018} by its internal properties. Then, Pak's scheme is degenerated into the version analyzed in \cite{Wang:Cryptanalysis:SP2018}. Based on the preliminary cryptanalysis works given in \cite{Wang:Cryptanalysis:SP2018} and \cite{Chen:Cryptanalysis:ND2018}, both the equivalent secret-key of Pak's scheme and all its unknown parameters are recovered by 
a little more pairs of chosen plain-image and the corresponding cipher-image in \cite{Zhucx:cryptanalysis:En2018}. In \cite{Zhu:analyze:IA2018}, Zhu et al. propose a chosen-plaintext attack on another one-round PTDWOS, where diffusion is implemented by modulo addition and XOR. The equivalent secret key of such diffusion mechanism can be easily recovered \cite{Zhang:diffusion:TC2018}.

In \cite{Lim:Cryptanalysis:IM2018}, M. Li et al. propose a chosen-plaintext attack on an image encryption scheme based on Ikeda map, where the Ikeda map is used to generate PRNS to control the mask operation XOR. In \cite{Lim:Cryptanalysis:SP2018}, M. Li et al. further present two chosen-plaintext attack methods on a colour image encryption using chaotic APFM (Amplitude Phase Frequency Model) nonlinear adaptive filter, which is composed of three parts: position permutation; nonlinear substitution; XOR operation.  In \cite{Lim:Cryptanalysis:SPIC2018}, M. Li et al. recover the permutation relation of image encryption scheme using PTDWOS with some chosen plain-images. In \cite{Lim:hybrid:IA2018}, M. Li et al. recover the equivalent secret key of an encryption algorithm using one-round PTDWOS with 512 pairs of chosen plain-images and the corresponding cipher-images, where hybrid hyper-chaos and CA are used as the  source of PRNS to control the permutation and mask operation XOR. In \cite{Lim:Cryptanalysis:MTA2018}, another encryption scheme using PTDWOS is cryptanalyzed. Similar to the strategy given in \cite{cqli:autoblock:IEEEM18}, the sensitivity mechanism of PRNG on the plain-image can be easily canceled by some special plain-images.

In \cite{Ahmad:Cryptanalysis:Symmetry2018}, security defects and chosen-plaintext attack on an image encryption schemes for body area network system are analyzed. Furthermore, some additional operations are adopted to withstand the proposed attack: using SHA-512 to build linkage between the plain-image and PRNS; updating initial conditions and control parameters of the chaotic system; adding more complex encryption operations. However, the incurred computational load is not considered for the required security level.

\subsection{Cryptanalysis of some image encryption schemes with complex structure}

Once the basic encryption operations are repeated some rounds or S-box is used, the structure of the whole encryption scheme become much more complex, which dramatically increases the hardship of the corresponding cryptanalysis.

In \cite{Diab:permutation:SP2018}, two chosen plain-images of fixed value are constructed to recover the equivalent secret key of an ``efficient image cryptographic algorithm'' reusing the permutation matrix generated by Cat map or Baker map dynamically. But, how to perform the attack on the PTDWOS with multiple rounds are not theoretically proved in \cite{Diab:permutation:SP2018}.

In \cite{Mousa:diffusion:MTA18}, some special plain-images are selected to narrow the scope of some subkeys of a PTDWOS with less than or equal to 3 rounds.
In \cite{Dhall:Cryptanalysis:SP2018}, the usability of a 4-round diffusion-then-permutation without substitution-box is questioned, which can be considered as a special form of 5-round PTDWOS. Furthermore, the linear relationship between plain-image and the corresponding cipher-image is derived as the adopted permutation is a static rotation of 90 degrees, which is used to support the proposed differential attack. Similar to the strategy introduced in \cite{Cqli:Fridrich:SP2017}, the permutation relation of a multiple-round PTDWOS
is recovered by checking how the changes of the cipher-pixels influence that of the plain-pixels in the scenario of chosen-ciphertext attack \cite{Hoang:analysis:O2018}. In \cite{Zhucx:analysis:Symmetry2018}, two chosen plain-images are constructed to recover an equivalent version of S-box and PRNS of a chaos-based image encryption scheme using S-box.

\subsection{Other image cryptanalysis}

According to the scenario or theory used by an encryption scheme, a special cryptanalysis method may work very well.

In \cite{Xiong:asymmetric:JOSAA2018}, a ciphertext-only attack on a double-image optical encryption technique based on an asymmetric (public-key encryption) algorithm is proposed using phase retrieval algorithm with median filtering and normalization operation. As for an optical cryptosystem based on phase-truncated Fourier
transform and nonlinear operations, the same research group presented a specific attack based on phase retrieval algorithm with normalization and a bilateral filter in \cite{Xiong:optical:OC2018}. In \cite{Singh:elliptic:O2018}, a joint compression and encryption scheme based on elliptic curve cryptography
(ECC) is cryptanalyzed. Due to the used parameter of ECC is too small, three known attacking methods on solving elliptic curve discrete logarithmic problem are used to recover the private key from the public key.

In \cite{Hai:Cryptanalysis:OPIP2018}, a large number of plaintexts and the corresponding ciphertexts encrypted by 
any optical encryption technique based on random phase encoding are used as
training samples for optimizing the parameters of a deep neural network, which can reconstruct any plain-image from its corresponding cipher-image. Actually, deep learning can also be used to test security of an encryption scheme via adversarial training.

In \cite{Luo:hardware:IJBC:2018}, Luo et al. evaluates the security of chaotic encryption schemes by using side-channel attack. Some intermediate values relating to the plaintext and the round keys can be revealed by observing the power consumption. In \cite{Annaby:circuits:MTA2018}, influences of different operations on resistance of chaotic image encryption schemes combining position permutation and a mask operation against chosen-plaintext attacks are discussed. Among some reversible and irreversible gates, including XOR, Toffoli and Fredkin gates, only Fredkin gate can help the supported encryption scheme to withstand chosen-ciphertext attack.

About eighty publications on ``reversible data hiding in the cipher-image (encrypted domain)'' were published in the past decade. Many of them use simple stream cipher XOR to generate cipher-images. Using the strong correlation among neighboring bits in every bit-plane in a natural image, a cipher-text only attack is proposed in \cite{Fouad:stream:SP2018} to recover the approximate version of the original plain-image with good visual quality.

In \cite{Lin:stream:ND2018}, a chosen-ciphertext attack on a chaotic stream cipher using self-synchronization and closed-loop feedback is proposed. Some parameters of the used 3-D discrete time chaotic system can be estimated. In \cite{Lin:stream:IJBC2018}, Lin et al. further derive coefficients of the nominal matrix of a stream cipher based on 8-D self-synchronous chaotic system in the scenarios of known-plaintext attack and chosen-ciphertext attack, respectively.

\section{The challenges on design and analysis of image encryption schemes}
\label{challenge}

Based on the above review on two sides of image cryptology, we summarize some critical challenges on design and security analysis of image encryption schemes as follows.
\begin{itemize}
\item No specific application scenario is targeted in the design process of most image encryption schemes. Under such precondition, the balancing point among computation load, the real required security level and economic cost for a potential attack cannot be considered at all. Only a few image encryption schemes are designed for solving 
security challenges in specific application scenarios: surveillance \cite{Muhammad:surveillance:TII2018}, IIoT (industrial Internet of Things) \cite{Hu:synchronization:TII2018}, social media \cite{zhoujt:privacy:TMCCA2018}.

\item Special formats of image data are not considered. 
A large number of image encryption schemes just treat plain-image as an ordinary textual data \cite{yuhai:adaptive:SP2018}. Due to the large size of conventional 
image files, compression and encryption should be combined well \cite{He:encryption:ITM2018,Li:joint:TM2018}. As \cite{Noura:medical:MTA2018}, only ROI (region of interest) of DICOM (Digital Imaging and Communications in Medicine) medical images are encrypted with one round of permutation and substitution to satisfy real-time processing requirement.


\item Importance of S-box is seriously neglected. As a lookup table, S-box is an
efficient way to obscure the relationship between the secret key and the cipher-image. Strangely enough, it is seldom adopted in image encryption schemes.

\item The underlying weak randomness of chaos-based PRNG is widely omitted.
The periods of PRNS obtained by iterating Logistic map and skew tent map in digital computer are rigorously analyzed in \cite{cqli:network:TCASI2019}. 
Much more theoretical results on dynamical properties of chaotic maps, e.g. 2D-LASM, H\'enon map, Lorenz system in limited-precision arithmetic domain should
be further investigated by extending the analysis in the field $\mathbb{Z}_2$ to $\mathbb{Z}_{2^e}$ using advanced theoretical analysis tools, e.g. Hensel's lifting lemma, where $e>1$.

\item The security test metrics are widely misused. As shown in \cite{Preishuber:motivation:TIFS2018,cqli:autoblock:IEEEM18}, the security 
assessment metrics used in most chaos-based image encryption schemes are unconvincing. So convincing methodologies for assessing security of image encryption schemes need to be established urgently.

\item The previous lessons drawn by tough cryptanlaysis are ignored by the designers of new image encryption schemes. For example, the sensitivity mechanisms of PRNS on the plain-image built in \cite{BenSlimane:DNA:MTA2018,Zhu:analyze:IA2018,Kumar:model:JISA2018,He:encryption:ITM2018} can be easily canceled by constructing some special plain-images, which has been reported in many cryptanalysis papers such as \cite{cqli:IEAIE:IE18,cqli:autoblock:IEEEM18}.

\item Prejudice on the value of cryptanalysis obstructs the progress of image security. As emphasized in the abstract of the special report on PRNG test suite of \href{https://www.nist.gov/publications/statistical-test-suite-random-and-pseudorandom-number-generators-cryptographic}{NIST}, ``statistical testing cannot serve as a substitute for cryptanalysis''. It is unreasonable to require a cryptanalyst to propose modifications to withstand the proposed attacks. The work of cryptanalysis should focus on the conditions of required plain-image; computational cost of the attack; space complexity. In 2018, improvement (enhancement, remedy) methods were proposed to fix the reported security defects in a number of papers on image cryptanalysis as references \cite{Zhucx:analysis:Symmetry2018,Diab:permutation:SP2018,Wu:permutation:SP2018,Lim:Cryptanalysis:IM2018,Wang:Cryptanalysis:SP2018,Panwar:Cryptanalysis:JEI2018,Chen:Cryptanalysis:ND2018,Zhucx:cryptanalysis:En2018,Feng:encode:IPJ2018,Diab:hyper:SP2018,Ahmad:Cryptanalysis:Symmetry2018,Zhu:analyze:IA2018,Annaby:circuits:MTA2018}. In fact, the proposed improvement methods fail to obtain the expected object due to the following subjective and objective reasons:
\begin{itemize}
\item The authors proposing a successful attack, e.g. chosen-plaintext attack in \cite{Diab:hyper:SP2018}, on a given image encryption scheme may not own comprehensive knowledge on designing secure image encryption schemes.

\item Many image encryption schemes are not designed and presented following the basic rules and general guidelines given in \cite{ozkaynak:review:ND18,Li:Rules:IJBC2006}, which makes essential improvement on the security level of the analyzed schemes very hard.
\end{itemize}
To cryptanalyze the seemingly complex image encryption scheme (e.g. multi-round PTDWOS), the modern advanced attacking methods in classic text cryptanalysis should be utilized, especially that published in the \href{http://faculty.cs.tamu.edu/guofei/sec_conf_stat.htm}{top security conferences} recently.
\end{itemize}

\section{Conclusion}

This paper reviewed the representative works on protecting
and attacking image data proposed in 2018 from the viewpoint of a senior image cryptanalyst. The representative works on the two sides were classified and some creative
ideas and methods were highlighted. More importantly,
the challenges on design and analysis of image encryption
schemes were briefly summarized to promote the
progress of protection level of image data in cyberspace.

\section*{Acknowledgement}

This research was supported by the Natural Science Foundation of China (No.~61772447, 61532020), the Joint Funds of the National Natural Science Foundation of China, China General Technology Research Institute (No.~U1736113) and the Fundamental Research Funds for the Central Universities (No.~531118010348).

\bibliographystyle{elsarticle-num}
\bibliography{BIB/image_secu_18,BIB/image_secu_18_analysis,BIB/image_secu_18_IEEE,BIB/image_secu_18_IA}

\end{document}